УДК 539.1.07

# УСТАНОВКА ЭПЕКУР ДЛЯ ПОИСКА УЗКИХ БАРИОННЫХ РЕЗОНАНСОВ В ПИОН-ПРОТОННОМ РАССЕЯНИИ

И. Г. Алексеев, И. Г. Бордюжин, П. Е. Будковский,
Д. В. Калинкин, В. П. Канавец, Л. И. Королева, А. А. Манаенкова,
Б. В. Морозов, В. М. Нестеров, В. В. Рыльцов, Д. Н. Свирида,
А. Д. Сулимов, Д. А. Федин

*Институт теоретической и экспериментальной физики, г. Москва, Россия*

В. А. Андреев, В. В. Голубев, А. Б. Гриднев, А. И. Ковалев,
Н. Г. Козленко, В. С. Козлов, А. Г. Кривший, Д. В. Новинский,
В. В. Сумачев, В. И. Тараканов, В. Ю. Траутман, Е. А. Филимонов

*Петербургский институт ядерной физики, г. Гатчина, Россия*

**М. Садлер (M. Sadler)**

*Абиленский христианский университет, Абилен, Техас, США*

*(Abilene Christian University, Abilene, Texas, USA)*

Экспериментальная установка ЭПЕКУР предназначена для поиска узких резонансных состояний путем прецизионного измерения дифференциальных и полных сечений реакций пион-нуклонного взаимодействия с шагом по энергии пионов 1 МэВ. За пять лет, прошедших от появления идеи эксперимента до начала набора статистики в апреле 2009 г., на универсальном пучковом канале 322 протонного синхротрона У-10 ИТЭФ создан безмагнитный спектрометр с жидководородной мишенью на основе широкоапертурных проволочных дрейфовых камер с гексагональной структурой. Уникальные свойства магнитного канала позволяют измерять импульс пучковых частиц с точностью лучше 0.1% с помощью пропорциональных камер с шагом проволочек 1 мм, расположенных в первом фокусе канала. Многочисленные подсистемы установки разработаны с использованием современной микроэлектроники, включая микропроцессоры и микросхемы программируемой логики, отлажены и испытаны как автономно, так и в составе установки. Распределенная система сбора и накопления информации базируется на распространенных интерфейсах USB и Ethernet, что позволяет использовать готовые решения для этих протоколов и обеспечить высокую производительность.

## ВВЕДЕНИЕ

Одной из загадок квантовой хромодинамики в настоящее время является отсутствие твердо установленных экзотических адронов, существование которых теория не запрещает. В 2003 году в ИТЭФ [1] и КЕК [2] был обнаружен узкий экзотический барион $\theta^+$ со странностью +1 и массой 1540 МэВ, ранее предсказанный в киральной солитонной модели [3]. В дальнейшем появилось довольно много работ в которых этот резонанс не наблюдался (см. обзор и ссылки в [4]). Позже авторы оригинальных работ повысили значимость своих результатов [5] и [6]. Коллаборация CLASS на большом статистическом материале смогла лишь оценить верхний предел сечения фоторождения $\theta^+$ [7]. Несколько позже группа той же коллаборации увидела уверенный сигнал этого резонанса, отбирая события в области интерференции с $\varphi$-мезоном [8]. Таким образом, вопрос о существовании $\theta^+$ остается открытым. Квантовые числа $\theta^+$ не определены экспериментально, но можно предполагать, что если он существует, то принадлежит к $½^+$ антидекуплету, предсказанному Д. Дьяконовым, В. Петровым и М. Поляковым в 1997 году [9]. В этом случае должен существовать также и нестранный P11 резонанс с массой около 1700 МэВ. Определенные указания на существование такого резонанса имеются в результатах модифицированного парциально-волнового анализа группы университета Джорджа Вашингтона [10] при массах 1680 и 1730 МэВ [11]. С другой стороны, в фотореакциях на дейтоне в эксперименте GRAAL наблюдается узкая особенность с массой 1685 МэВ и шириной менее 30 МэВ как при рождении η-мезона [12], так и в комптоновском рассеянии [13]. Аналогичная особенность видна и в более поздних работах по фоторождению на нейтроне [14,15]. Основной идеей, вдохновившей появление данной работы, стала беспрецендентная возможность поиска и исследования нестранного нейтрального члена антидекуплета пентакварков на пионных пучках синхротрона У-10 ИТЭФ. Такую задачу можно решить, непосредственно наблюдая проявления резонанса в дифференциальных и полных сечениях реакций пион-нуклонного взаимодействия, при условии достижения высокой точности результата и при достаточно мелком шаге по энергии. Исключительность предложенного подхода определяется одновременно несколькими факторами. Прочие указания на существование подобного состояния получены с помощью электромагнитных пробников, тогда как попыток исследовать вопрос при помощи адронов не предпринималось, вероятно, в связи с отсутствием на тот момент пионных пучков подходящих энергий на других ускорителях мира. В отличие от подавляющего большинства экспериментов, в которых поиск пентакварков производится в множественном рождении ("production"), предложенный подход предполагает рождение резонанса непосредственно в s-канале ("formation"). Единственной сигнатурой исследуемого криптоэкзотического члена антидекуплета является его малая ширина, предсказанная авторами [9]. Прецизионное, лучше 0.1%, измерение импульса каждой частицы пучка, возможное благодаря уникальным свойствам пучковых каналов ИТЭФ, может позволить пронаблюдать очень узкую, шириной в несколько МэВ, особенность в сечении. Абсолютная калибровка центрального импульса пучка с точностью лучше 0.1% задает точность определения массы резонанса на уровне 0.5 МэВ. Авторами идеи пентакварков предполагается незначительная связь изучаемого резонанса с $\pi p$-каналом; однако, исследуя поведение парциально-волновых анализов, можно показать, что его проявления в дифференциальном сечении упругого $\pi p$-рассеяния можно надежно заметить даже если относительный бранчинг в пион-нуклонный канал составляет всего 5%. Наконец, прецизионное измерение дифференциальных сечений позволит выполнить парциально-волновой анализ на новом качественном уровне и однозначно определить все квантовые числа исследуемой частицы.

## УНИВЕРСАЛЬНЫЙ МАГНИТНО-ОПТИЧЕСКИЙ КАНАЛ

Магнитно-оптический канал 322 (рис. 1) спроектирован специально для экспериментов с точным измерением импульса частиц пучка [16] и построен по двухступенчатой ахроматической схеме. Внутренняя мишень Т303 расположена в вакуумной камере основного кольца протонного синхротрона У-10. Мишень изготавливается из бериллиевой фольги толщиной 100-200 мкм в форме треугольного флажка, острый носик которого вводится в пучок протонов основного кольца по окончании фазы ускорения 4-х секундного цикла синхротрона на 500–800 мс. Область взаимодействия имеет размер $d_\text{М}$≈2–3 мм.

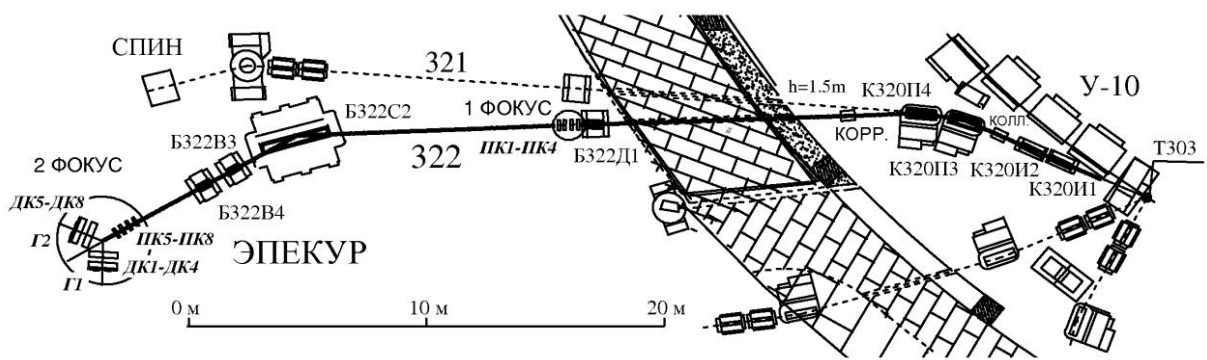

**Рис. 1.** *План-схема универсального магнитно-оптического канала 322 протонного синхротрона У-10 ИТЭФ*

Рожденные в мишени частицы захватываются в канал под углом 10.5° к оси пучка ускорителя при помощи магнитных линз К320И1 и К320И2, которые собирают вторичный пучок в первом фокусе канала на расстоянии 26 м от мишени. Дипольные магниты К320П3 и К320П4 поворачивают пучок на угол 40° и обеспечивают спектрометрические возможности канала. Частицы с импульсом, отличающимся на величину $\Delta P$ от среднего импульса канала, оказываются сфокусированы на расстоянии $\Delta X_1 = D_1 \cdot \Delta P$ от геометрической оси канала по горизонтали ($D_1$=54 мм/%– дисперсия в первом фокусе), что позволяет получать точное значение импульса каждой частицы, измеряя координату пересечения ее трека с фокальной плоскостью. Точность такого измерения, или импульсное разрешение канала, определяется многими факторами, включая качество изготовления и установки магнитных элементов и стабильность поддержания их токов, однако, в основном, эта величина определяется размером области взаимодействия на внутренней мишени и может быть оценена как $\sigma_P/P = R_1 \cdot d_M/D_1 \approx 0.05$–$0.08\%$ ($R_1$=1.47 – оптическое увеличение мишени в первом фокусе в горизонтальной проекции). Пара линз Б322В3 и Б322В4 создают во втором фокусе перевернутое изображение образа мишени, сформированного в первом фокусе канала, с увеличением $R_2$=0.76, а магнит Б322С2 обеспечивает поворот пучка и дисперсию при переходе из первого во второй фокус $D_2$=41 мм/%. Полная ахроматичность пучка достигается при выполнении соотношения $D_2/R_2 = D_1$. Действительно, для координаты частицы во втором фокусе можно написать $\Delta X_2 = -R_2 \cdot \Delta X_1 + D_2 \cdot \Delta P = (-R_2 \cdot D_1 + D_2) \cdot \Delta P$; ахроматичность достигается при независимости этой координаты от импульса частицы, т.е. при $-R_2 \cdot D_1 + D_2 = 0$.

Для экспериментальной проверки импульсного разрешения канала исследовались импульсные спектры рассеянных протонов в реакции p+$^9$Be→p+$^9$Be$^*$. Монохроматические протоны с импульсом 1.5 ГэВ/с, циркулирующие в кольце ускорителя, испытывали взаимодействие на внутренней бериллиевой мишени. Протоны, рассеянные на угол 10.5° упруго и с возбуждением нижних уровней $^9$Be, захватывались магнитно-оптическим каналом, а их импульсный спектр анализировался при помощи трех пропорциональных камер ПК1-ПК3, расположенных в разрыве вакуумной трубы в первом фокусе, путем измерения горизонтальной координаты пересечения трека с фокальной плоскостью. Как следует из работ [17] и [18], в таких кинематических условиях наряду с упругим рассеянием интенсивно происходит рассеяние с возбуждением уровня ядра бериллия 2.43 МэВ с шириной 0.77 кэВ и двух близких уровней 6.37 и 6.76 МэВ с суммарной шириной около 1.35 МэВ. В импульсном спектре указанным линиям должны соответствовать пики, отстоящие от пика упругого рассеяния на 2.87 и 7.97 МэВ/с. На рис. 2 представлен экспериментально полученный импульсный спектр, отфитированный тремя гауссовыми распределениями на фоновой подложке.

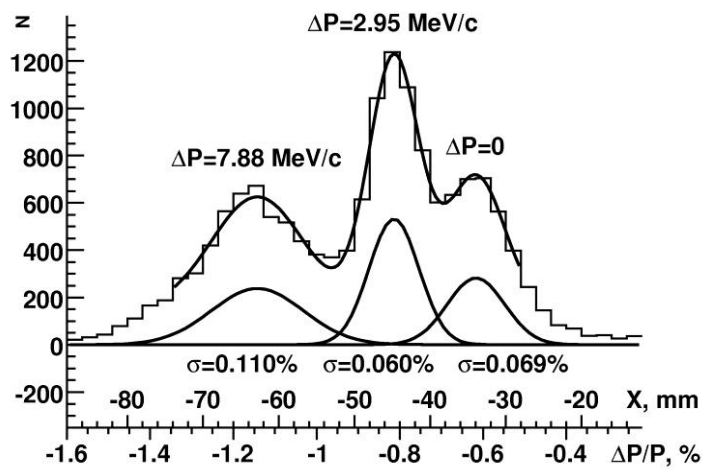

**Рис. 2**. *Распределение частиц пучка по горизонтальной координате в фокальной плоскости первого фокуса при настройке канала на реакцию $p+^9Be \rightarrow p+^9Be^*$ при импульсе протонов 1.50 ГэВ/с; нижняя шкала соответствует отличию импульса частиц от центрального импульса канала в процентах*

Частицы в правой части распределения, соответствующие импульсам, большим, чем при упругом рассеянии, практически отсутствуют. Положения пиков хорошо соответствуют предсказанным. Ширина пика при 7.88 МэВ/с определяется, в основном, естественной шириной уровня, тогда как для упругого рассеяния и пика первого возбужденного состояния основной вклад в ширину вносит импульсное разрешение канала, которое может быть оценено величиной 0.06–0.07%.

Представленное распределение получено при специальной настройке ускорителя – улучшенная монохроматичность внутреннего пучка, его наводка на самый конец флажка мишени. В реальных условиях работы, оптимизированных на максимальную интенсивность пучка, импульсное разрешение составляет 0.12–0.15%, прежде всего из-за увеличения размера области взаимодействия пучка ускорителя с флажком внутренней мишени. При оптимальной настройке достигаются следующие параметры пучка: размер пятна во втором фокусе 5.5 мм по горизонтали и 3.5 мм по вертикали, угловая сходимость 5 мрад и 10 мрад, соответственно, длительность сброса 500–800 мс каждые 4 с, интенсивность π-мезонов при импульсе в ускорителе 9 ГэВ/с 150–250 тыс. частиц за сброс.

Использование реакции p+$^9$Be→p+$^9$Be$^*$ на внутренней мишени ускорителя совместно с прецизионным измерением магнитного поля поворотного магнита Б322С2 с помощью ЯМР позволяет выполнить калибровку абсолютной величины среднего импульса канала с точностью лучше, чем 0.1%. Импульс протонов в кольце У-10 устанавливается близким к требуемому калибровочному значению и измеряется фактическое значение ускоряющей частоты $f$. Точное значение импульса протонов $p$ в выбранном пике возбуждения определяется по формулам:

$$p = p_0 - \frac{p_0^2(1-\cos\alpha) + m_{Be} \cdot \Delta m_{Be}}{p_0 \cdot ((1-\cos\alpha) + m_{Be}/\sqrt{m_p^2 + p_0^2})}; \quad p_0 = \frac{m_p \beta}{\sqrt{1-\beta^2}}; \quad \beta = \frac{L}{ct} = \frac{Lf}{nc},$$

где $p_0$, $\beta$ – импульс и скорость протонов в ускорителе, α=10.5º – угол захвата канала, $L$ – длина орбиты синхротрона, $n$ – кратность ускоряющей частоты $f$, $\Delta m_{Be}$ – изменение массы ядра бериллия за счет возбуждения. Магнитный канал приблизительно настраивается на вычисленный импульс $p$ на основе расчетных данных; перед настройкой токи всех элементов поднимаются до фиксированных максимальных значений для устранения гистерезисных явлений. Подбираются токи магнитов К320П3 и К320П4 таким образом, чтобы положение выбранного пика возбуждения в первом фокусе канала совпадало с геометрической осью пучка, т.е. производится точная настройка угла поворота для центрального импульса пучка $p$. Далее выполняется уточняющая оптимизация фокусирующих элементов первой ступени канала К320И1 и К320И2, и повторная подстройка поворотных магнитов для компенсации возможной дипольной компоненты в линзах (на практике не требуется). Затем в аналогичном порядке осуществляется настройка угла поворота и фокусировка второй ступени с целью наводки пучка на центр жидководородной мишени установки. При полностью настроенном канале с помощью ЯМР фиксируется значение магнитного поля поворотного магнита Б322С2, которое направляет пучок с центральным импульсом $p$ на мишень установки. Изложенная процедура калибровки выполнена при трех импульсах в рабочем диапазоне эксперимента и полученные значения поля Б322С2 и фокусирующих токов описаны линейной зависимостью. Установка произвольного требуемого импульса канала производится в обратном порядке: устанавливаются рассчитанное значение поля Б322С2 по ЯМР и токов линз, а затем подбором тока поворотных магнитов первой ступени достигается наводка на мишень установки, то есть выбирается как раз такой импульс частиц на геометрической оси первого фокуса, который доставляется на мишень заданным полем Б3Q2С2. Проверка калибровки в еще двух точках рабочего диапазона показала, что описанный метод обеспечивает точность установки центрального импульса канала не хуже 0.1%.

## УСТАНОВКА ЭПЕКУР

Установка ЭПЕКУР представляет собой двухплечевой безмагнитный спектрометр на основе пропорциональных и дрейфовых камер (рис. 3) и предназначена для регистрации и восстановления кинематики реакций упругого рассеяния заряженных пионов на протонах. Блок из 4 двухкоординатных пропорциональных камер ПК1-4 расположен в первом фокусе магнитно-оптического пучкового канала и обеспечивает определение импульса индивидуальной частицы пучка путем нахождения координаты пересечения ее трека с фокальной плоскостью. Аналогичный блок ПК5-8 расположен непосредственно перед жидководородной мишенью ЖВМ и позволяет восстанавливать треки налетающих пионов до вершины взаимодействия с ядрами водорода. Плечи спектрометра, образованные блоками широкоапертурных проволочных дрейфовых камер ДК1-4 и ДК5-8, расположены симметрично относительно оси установки. Каждый блок включает в себя по две однокоординатных камеры для определения вертикальных (ДК1, ДК3, ДК5, ДК7) и горизонтальных (остальные камеры) проекций треков рассеянного пиона и протона отдачи. Каждая камера имеет две сигнальные плоскости, так что восстановление треков вторичных частиц производится, в большинстве случаев, по 4 срабатываниям на треке в каждой из проекций. Расположение камер в блоках оптимизировано для регистрации треков рассеянных пионов в диапазоне 17-100$^\circ$ в лабораторной системе. Сцинтилляционные годоскопы Г1 и Г2, полностью перекрывающие чувствительные области последних камер обоих плеч, были задуманы для выработки триггера на события упругого рассеяния. Однако для устранения влияния эффективности триггера на угловую зависимость измеряемых сечений было принято решение не использовать годоскопы в триггерной логике. В результате для получения триггерного сигнала используются сцинтилляционные счетчики С1 и С2, вето-счетчик А1 и импульсы срабатывания пропорциональных камер ПК1-8, а годоскопические счетчики применяются только для анализа эффективности дрейфовых камер по всему размеру чувствительной зоны. Выделение событий упругого рассеяния производится по угловым корреляциям треков вторичных частиц, зарегистрированных в противоположных плечах спектрометра.

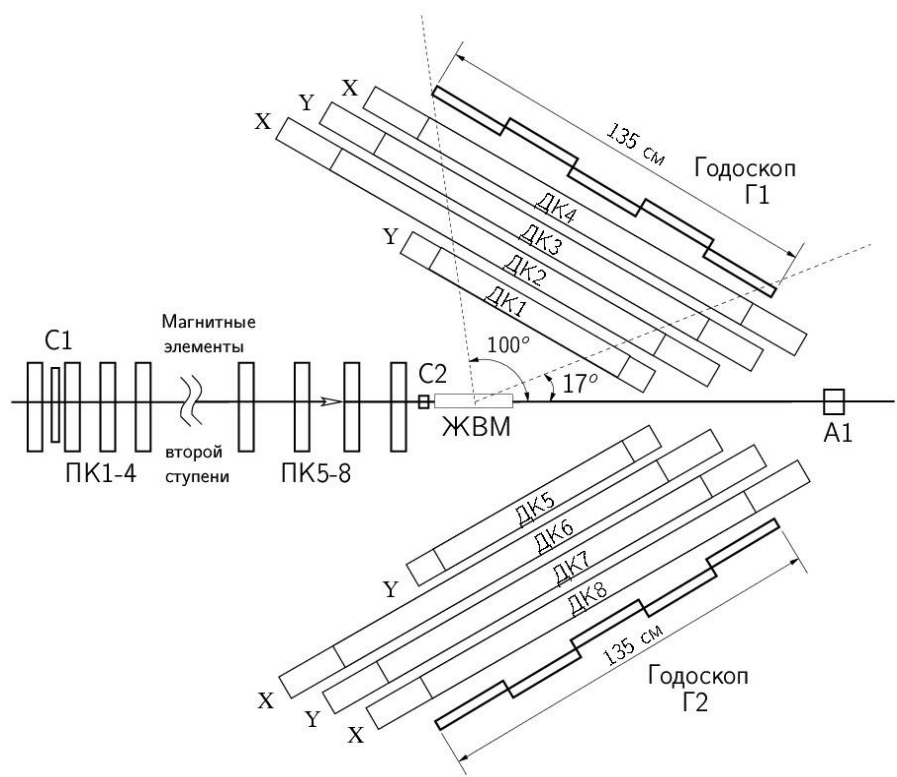

**Рис. 3**. *Схема установки ЭПЕКУР, вид сверху*

# ЖИДКОВОДОРОДНАЯ МИШЕНЬ

Водородная мишень установки представляет собой цилиндрический контейнер, заполненный жидким водородом и помещенный в вакуумный кожух для уменьшения теплопритока (рис. 4). Мишень вместе с ее системой поддержания режима работы является модификацией аппаратуры, использованной в [19,20]. Основной задачей модификации является уменьшение многократного рассеяния вторичных частиц на конструктивных элементах мишени и увеличение надежности работы за счет применения современных технологий. Цилиндрическая часть контейнера 1 диаметром 40 мм и донца 2 выполнены из майлара толщиной 100 мкм и закреплены на корпусе мишени 3 и тонком алюминиевом кольце 4 с помощью эпоксидного клея; места крепления усилены намоткой капроновой нити, также пропитанной эпоксидным клеем. Длина цилиндрической части мишени 250 мм, полная длина по пучку около 290 мм. Жидкий водород из камеры ожижения заполняет объем контейнера по трубке 8, а возврат газообразной фазы происходит по трубке 9 из верхней части корпуса мишени 3. Бериллиевый стакан 5 диаметром 80 мм с толщиной стенки 1 мм изготовлен из конструкционного бериллия плотностью 1,8 г/см$^3$ по специальному заказу на НПП

«МИКРОПРОМ» и прикреплен к корпусу вакуумного кожуха 6 с помощью прижимного фланца. Прозрачная мембрана 7 (майлар 100 мкм) обеспечивает минимальное взаимодействие налетающих частиц пучка и возможность визуального контроля уровня водорода в мишени. Для визуального наблюдения используется зеркало из натянутого алюминизированного майлара, установленное под углом близким к 45° на траектории пучка. В рабочих условиях возможно дистанционное наблюдение при помощи USB-камеры, изображение с которой передается на экран оператора. Механическая фиксация мишени обеспечивается телескопической трубой 10, она же используется для создания изоляционного вакуума в кожухе мишени. В раздвинутом состоянии трубы 10 открывается доступ к местам соединений трубок 8, 9 с отводными трубками теплообменника.

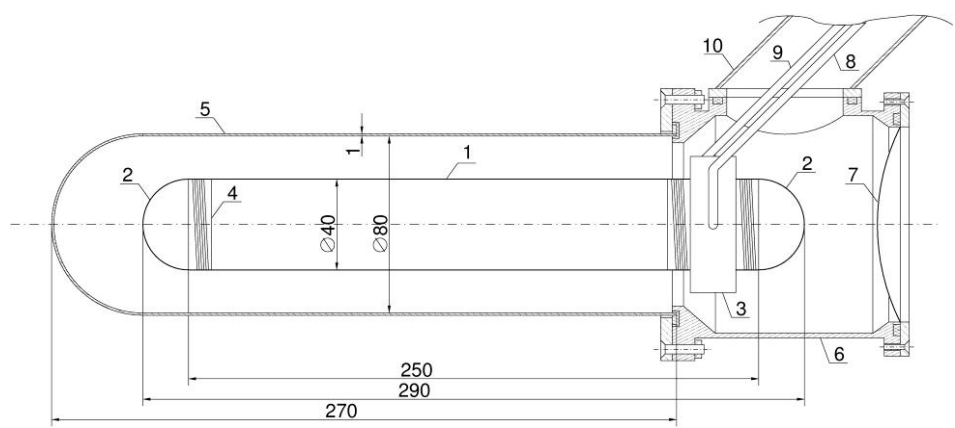

**Рис. 4**. *Разрез вакуумного кожуха жидководородной мишени*

Ожижение водорода достигается путем циркуляции холодного газообразного гелия через конденсор камеры ожижения КО (рис. 5). Жидкий водород, накапливающийся в нижней части камеры, заполняет контейнер мишени М под действием силы тяжести, а пары кипящего в нем водорода поступают в камеру ожижения на реконденсацию. Заполнение мишени и поддержание температурного режима ожижения осуществляется системой, схематично изображенной на рис. 5. Перед началом работы резервуар объемом 1.1 м$^3$ заполняется водородом из баллона через вентили В1 и В4 в количестве, необходимом для дальнейшего ожижения (приблизительно до 1.3−1.4 атм абсолютного давления). При этом вентили В2 и В3 закрыты, а водородный объем мишени откачан. Первичное накопление жидкого водорода осуществляется при открытых В1 и В3, при этом газ, поступающий из резервуара, получает предварительное охлаждение в наклонной штанге в водородно-гелиевом теплообменнике. Количество ожиженного водорода определяется по уменьшению давления в резервуаре с помощью мановакууметра МВ2. По заполнении контейнера мишени система переводится в режим стабилизации давления водорода с целью поддержания постоянной плотности и уровня жидкого водорода в контейнере. Режим стабилизации осуществляется с помощью специального клапана К1, регулирующего поток холодного гелия через коненсор. При недостаточном холодопритоке скорость испарения водорода превышает скорость конденсации, давление в системе становится выше величины опорного давления P1≈800 мм. рт. ст., клапан К1 открывается и поток гелия увеличивается. При избыточном охлаждении и падении давления ниже P1 клапан К1, наоборот, ограничивает скорость потока. В рабочем режиме вентиль В1 закрывают, сильно ограничивая водородный объем для увеличения чувствительности давления к количеству газообразной фазы, а трубка большего диаметра с открытым В2 обеспечивает быструю передачу давления из теплообменника на чувствительную мембрану К1. Контроль рабочего давления производится мановакууметром МВ1, а расход гелия измеряется поплавковым ротаметром РТ1. Избыточное давление в дьюаре создается за счет естественного испарения гелия в сосуде, а для его поддержания в заданных пределах используются клапаны К2 и К3. При превышении опорного давления P3=1.3 атм избыток давления сбрасывается через клапан К3 и контрольный ротаметр РТ2. В случае падения давления в дьюаре ниже P2=1.2 атм открывающийся клапан К2 обеспечивает добавку гелия из баллона. Величина давления подпора контролируется мановакууметром МВ3. Откачка вакуумного объема производится двухступенчатой системой «Pfeiffer Vacuum» HiCube 80 Eco [21], состоящей из турбомолекулярного насоса ТМН и форвакуумного насоса МН мембранного типа; остаточное давление в охлажденной системе около $10^{-7}$–$10^{-6}$ мбар. Аварийный пружинный клапан К4, настроенный на избыточное давление 0.4 атм, гарантирует сброс водорода в резервуар при неконтролируемом росте давления. В случае резкого возрастания давления и разрыва контейнера мишени водород, попавший в вакуумный объем системы, сбрасывается в атмосферу через клапан К5 с большим проходным отверстием, одновременно закрывается заслонка на входе турбомолекулярного насоса для предотвращения выхода его из строя (на рис. 5 не показана). На схеме также не приведены системы промывки водородного объема азотом и его вакуумной откачки, очистки рабочего газа в азотных ловушках и некоторые другие.

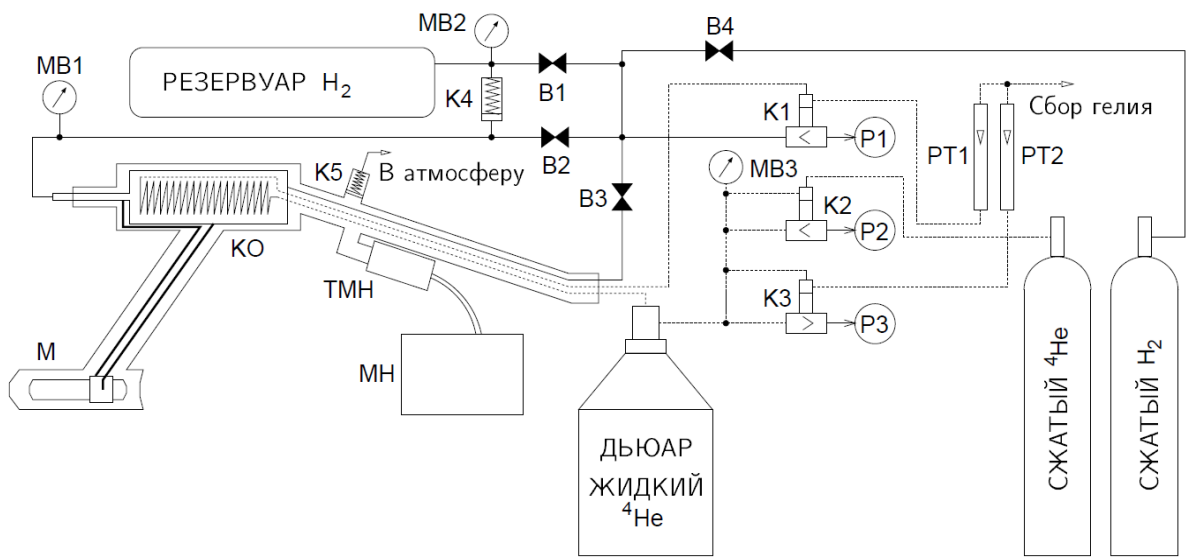

**Рис. 5.** *Упрощенная схема поддержания режима ожижения водорода ЖВМ. Сплошными линиями обозначены магистрали водорода, точечными – гелия.*

Система дистанционного контроля параметров мишени построена на базе модуля оцифровки E14-140-M фирмы «L-Card» [22]. Контроль давлений производятся датчиками абсолютного давления «Honeywell» SPT4V0030PA5W02 [23], установленными параллельно мановакуумметрам в ключевых точках системы. Широкодиапазонные измерители вакуума «Pfeiffer Vacuum» PKR 251 [21] установлены вблизи теплообменника и на входе турбомолекулярного насоса и формируют сигнал для оцифровки совместно с контроллером TPG 262 той же фирмы [21]. Температурные параметры системы измеряются в 7 точках криогенной части системы при помощи объемных резисторов ТВО-0,25 номиналом 2.1 кОм при комнатной температуре. Измерительный ток 0.75 мА поочередно подается на резисторы короткими импульсами с помощью прецизионного коммутируемого источника тока, управляемого логическими сигналами E14-140-M, так что средняя мощность, выделяемая в резисторах при измерении не превосходит 0.5 мВт. Приложение «LabView» позволяет видеть и запоминать историю изменения параметров, а также передавать значения параметров в центральный компьютер системы сбора данных. Запись значений давления в водородном объеме совместно с набираемой статистикой дает возможность производить вычисления плотности жидкого водорода в мишени в соответствии с [24].

Жидководородная мишень ЭПЕКУР надежна в эксплуатации. Время первичного ожижения водорода составляет 3-4 часа. Расход жидкого гелия в режиме стабилизации около 2 л/час при точности поддержания рабочего давления не хуже 1 мм. рт. ст., потери времени на замену дьюара 20-30 минут в сутки.

## ПРОПОРЦИОНАЛЬНЫЕ КАМЕРЫ

Пропорциональные камеры используются для определения импульса частиц пучка по траектории в первом фокусе магнитного канала (ПК1–ПК4), а также для нахождения параметров треков частиц перед взаимодействием в жидководородной мишени (ПК5–ПК8). Требования по точности определения координат привели к необходимости изготовления камер с шагом сигнальных проволочек 1 мм. Вопреки распространенному мнению о сложности работы с камерами с таким шагом, наши камеры показали надежную работу с высокой эффективностью при загрузках до $2 \cdot 10^6$ частиц в секунду.

Камеры разработаны и изготовлены в ПИЯФ, г. Гатчина, и по конструкции несильно отличаются от традиционных вариантов [25-27]. Каждый модуль камер (рис. 6) включает в себя две плоскости сигнальных проволочек 3 и 6, натянутых во взаимно-перпендикулярных направлениях для определения одновременно двух координат. Диаметр проволочек 15 мкм. Чувствительная область имеет размеры 200x200 мм$^2$. В отличие от камер [25-27] с проволочными потенциальными плоскостями, для уменьшения количества вещества на пути пучка высоковольтные электроды 2, 4, 5 и 7 выполнены из холоднокатаной фольги из сплава АБ2 толщиной 40 мкм. Зазор между сигнальной плоскостью и высоковольтными электродами 3 мм. Высокое напряжение подается на катодные плоскости независимо для каждой половинки камеры. Газовый объем камеры – общий для обеих половинок и ограничен майларовыми окнами 1 и 8 толщиной 50 мкм. Продув камер осуществляется газовой смесью 74.85% Ar, 25% изобутан, 0.15% CF$_3$Br, в которой по сравнению с т. н. "магической" количество фреона-13В1 сокращено вдвое.

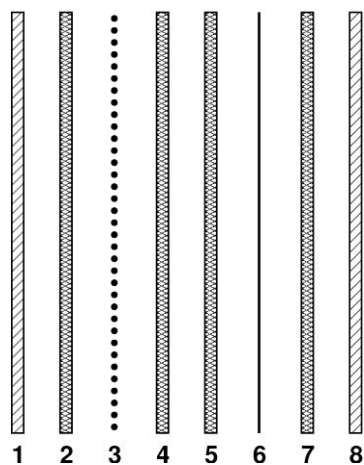

**Рис. 6**. Устройство двухкоординатных пропорциональных камер ЭПЕКУР с шагом 1 мм (не в масштабе): 3,6 – плоскости сигнальных проволочек, 2,4,5,7 – высоковольтные электроды, 1,8 – майларовые окна, ограничивающие газовый объем

Для регистрации сигналов срабатывания проволочек в ИТЭФ разработаны специальные платы электроники, каждая из которых включает в себя 100 каналов усилителей и дискриминаторов, а также цифровую часть, которая обеспечивает кодирование номеров сработавших проволочек и формирование и передачу пакетов данных по интерфейсу USB2.0. Предусмотрено также формирование быстрого сигнала "ИЛИ всех сработавших каналов", который используется для выработки триггера установки. Электроника считывания расположена непосредственно на рамках модулей камер; для полного обслуживания одного модуля требуется 4 100-канальных платы. Более подробное описание электроники считывания можно найти в [28].

В результате многочисленных тестов камеры показали стабильную работу в условиях различных загрузок вплоть до $4 \cdot 10^6$ частиц в секунду. На рис. 7 приведены типичные кривые зависимости эффективности камеры от напряжения на высоковольтном электроде при различных порогах дискриминации. Плато характеристики составляет 150–200 В, эффективность на плато – более 99%.

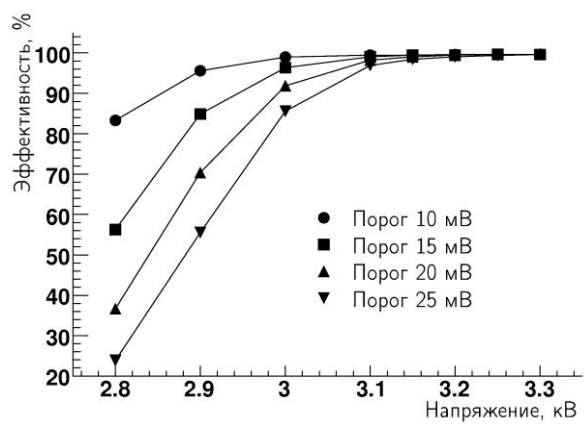

**Рис. 7**. *Зависимость эффективности пропорциональной камеры от напряжения на потенциальных плоскостях при различных порогах дискриминации*

Методический интерес может представлять вопрос о распределении времен прихода сигналов от сработавших проволочек. Электроника считывания позволяет программно перестраивать ширину окна, в котором регистрируются срабатывания, а также его положение относительно сигнала триггера установки. На рис. 8 приведена зависимость эффективности регистрации частицы от задержки чувствительного окна для треков, перпендикулярных плоскости камеры, ширина окна и шаг сканирования составляют 10 нс. Наиболее вероятный момент регистрации определяется разбросом времен дрейфа ближайшего кластера заряда на треке и отстоит приблизительно на 20 нс от момента прохождения частицы. Максимальное расстояние от трека частицы до проволочки 0.5 мм. Правое «плечо», предположительно, соответствует относительно более редким случаям разделения первичного заряда трека между соседними проволочками, в том числе за счет образования δ-электронов. В этих случаях возможно попадание заряда в поле проволочки, например, вблизи потенциального электрода, то есть на расстоянии до 3 мм от нее. Времена дрейфа в подобных ситуациях могут превосходить 100 нс. В рабочем режиме ширина окна чувствительности выбирается равной 80 нс, его положение показано на рис. 8 точечными линиями, а эффективность регистрации близка к 100%. Игнорирование сигналов с большими задержками за правой границей окна практически не сказывается на уменьшении эффективности регистрации, поскольку заряд от треков, дающих поздний сигнал на данной проволочке, сосредоточен, в основном, в области поля соседней проволочки и обеспечивает ее надежное срабатывание.

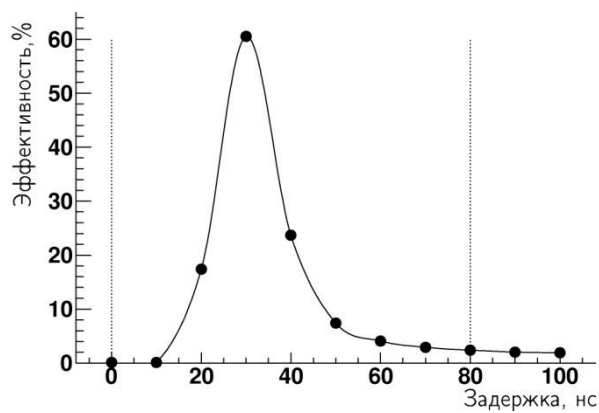

**Рис. 8**. *Зависимость эффективности пропорциональной камеры от задержки окна чувствительности по отношению к триггеру установки. Ширина окна 10 нс. Сдвиг временной шкалы выбран произвольно. Точечные линии отмечают настройку окна чувствительности в рабочем режиме.*

## ДРЕЙФОВЫЕ КАМЕРЫ

Дрейфовые камеры используются для регистрации треков вторичных частиц после взаимодействия пионов в жидководородной мишени. Для уменьшения фона при выделении событий упругого рассеяния по угловым корреляциям необходимо обеспечить хорошую точность определения углов треков, в частности минимизировать многократное рассеяние вторичных частиц. Из возможных конструкций дрейфовых камер была выбрана та, которая обеспечивает минимальное количество вещества на пути регистрируемых продуктов реакции, а именно -- проволочная дрейфовая камера с гексагональной структурой. Конфигурация почти правильных шестигранных дрейфовых ячеек иллюстрируется рис. 9. Открытыми кружками изображены потенциальные проволочки диаметром 100 мкм. Проволочки в слоях 3,4,6,7,9,10 находятся под высоким положительным потенциалом, образуя почти цилиндрически симметричное дрейфовое поле вокруг заземленных сигнальных проволочек (изображены черными точками в слоях 5 и 8). Диаметр сигнальных проволочек 25 мкм. Потенциальные плоскости 2 и 11 имеют нулевой потенциал и служат для симметризации поля ячеек с внешней стороны. Действительно, по отношению к верхнему (по рисунку) ряду ячеек проволочки плоскости 2 задают почти такую же конфигурацию потенциала, как и сигнальные проволочки плоскости 8. Шаг ячеек структуры 17 мм. Сдвиг ячеек в двух сигнальных плоскостях на половину шага позволяет эффективно определять, с какой стороны от сигнальной проволочки прошел трек регистрируемой частицы.

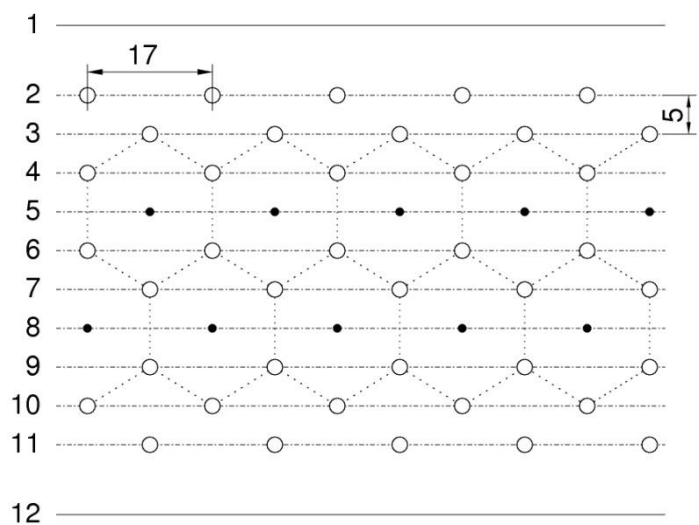

**Рис. 9**. *Часть периодической структуры проволочных дрейфовых камер ЭПЕКУР (не в масштабе): 2,3,4,6,7,9,10,11 – плоскости потенциальных проволочек, 5,8 – плоскости сигнальных проволочек, 1,12 – майларовые окна, ограничивающие газовый объем*

Дрейфовые камеры также сконструированы и изготовлены в ПИЯФ по аналогии с [29]. У больших камер ДК2-4 и ДК6-8 (рис. 3) чувствительная область имеет размер приблизительно 800х1200 мм$^2$, у меньших камер ДК1 и ДК5 – 400х600 мм$^2$. Рамы камер изготовлены из квадратного в сечении алюминиевого профиля «8» фирмы Item Profile [30], сторона квадрата 80 мм. Натяжение проволок регулируется микрометрическими винтами. Газовый объем камер ограничен майларовыми окнами 1, 12 (рис. 9) с герметизацией по алюминиевой раме. Рабочая газовая смесь содержит 70% аргона и 30% углекислого газа. Предусмотрена возможность подмешивания до 3% фреона-14 ($CF_4$) для очищения элементов внутреннего объема. Камеры ДК2, ДК4, ДК6 и ДК8 с расположением проволок вдоль короткой стороны имеют 72 ячейки в каждой сигнальной плоскости. В остальных камерах проволочки натянуты вдоль длинной стороны и образуют по 48 ячеек в плоскости у ДК3 и ДК5 и по 24 ячейки у меньших ДК1 и ДК5. Платы электроники считывания [28] расположены на рамах камер. Одна плата обслуживает 24 дрейфовые ячейки из одной сигнальной плоскости и обеспечивает усиление сигналов, их дискриминацию с общим программно-регулируемым порогом, определение времени прихода каждого сигнала с эффективной частотой дискретизации 576 МГц и формирование блока данных для передачи по интерфейсу USB2.0.

Во время испытаний и набора статистики эксперимента камеры показали стабильную и надежную работу в широком диапазоне питающих напряжений и различных составах газовой смеси. На рис. 10 приведена зависимость эффективности камеры от высокого напряжения, поданного на потенциальные проволочки гексагональной структуры, для газовой смеси 65% Ar и 35% $CO_2$. Из рисунка видно, что в диапазоне рабочих порогов 15-20 мВ ширина плато эффективности составляет 200−300 В. При продуве камеры рабочей газовой смесью напряжения, при которых достигается такая же эффективность, уменьшаются приблизительно на 50 В.

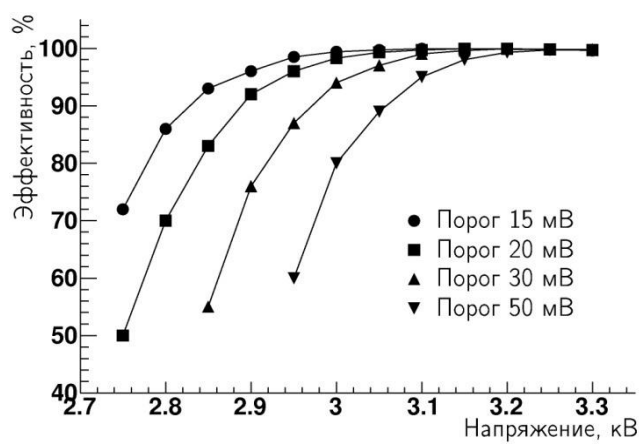

**Рис. 10**. *Зависимость эффективности дрейфовой камеры от питающего напряжения при различных порогах дискриминации*

Вычисление расстояния от сигнальной проволочки, на котором прошел трек заряженной частицы, производится на основании измеренного времени дрейфа заряда, порожденного этим треком. На рис. 11а представлено экспериментальное распределение $N(t)$ событий по времени дрейфа для случая, когда через камеру проходит широкий дефокусированный пучок перпендикулярно ее плоскости. В этих условиях можно считать, что треки равномерно распределены по отрезку $[0, R_0]$ между сигнальной проволочкой и границей ячейки. Тогда плотность вероятности найти трек на расстоянии $x$ от сигнальной проволочки $P_1(x)=p_0$ не зависит от $x$, причем $p_0 R_0 = 1$. Пусть теперь $x(t)$ есть искомая функция преобразования времени дрейфа $t$ в расстояние $x$, тогда $P_2(t)=P_1(x) \cdot x'(t)$ есть плотность вероятности найти событие с временем дрейфа $t$ (по свойству плотности вероятности), и $x'(t)=R_0 \cdot P_2(t)$. Заметим теперь, что распределение $N(t)$ рис. 11а, нормированное таким образом, чтобы его интеграл был равен единице, является экспериментальным приближением для функции $P_2(t)$. Тогда $x(t)$ может быть определена из данных следующим образом:

$$x(t) = \frac{R_0}{N_0} \int_{t_0}^{t_0+t} N(z)dz ,$$

где $N_0$ – полное число событий в распределении $N(t)$, а $t_0$ – время, в которое начинается подъем распределения $N(t)$, то есть время появления самых ранних сигналов ($t_0$=0 на рис. 11а). Полученная таким образом функция $x(t)$ приведена на рис. 11б. В реальных условиях возможно появление небольшой равномерной подложки у распределения $N(t)$ из-за случайных совпадений при больших загрузках или шумовых срабатываний при большом газовом усилении. В этих случаях применяется вычитание подложки путем анализа уровня распределения, непосредственно предшествующего времени $t_0$.

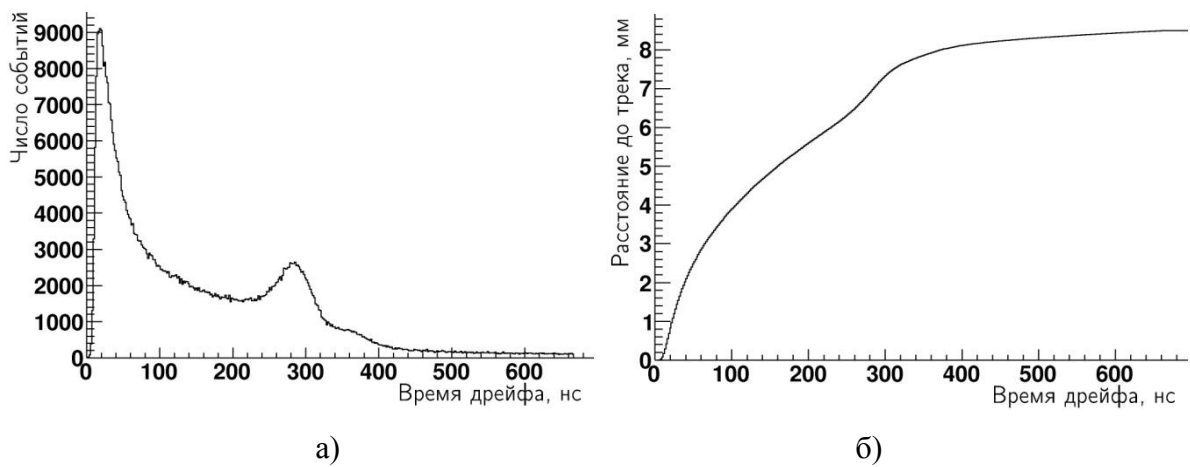

а)  б)

**Рис. 11**. *Распределение N(t) событий по временам дрейфа при равномерном распределении треков по ширине ячейки (а) и полученная из этого распределения функция x(t) преобразования времени дрейфа в расстояние от трека до сигнальной проволочки (б)*

При определении $x(t)$ из набранных экспериментальных данных используются только те ячейки камер, для которых треки рассеянных частиц пересекают плоскость камер под углами, близкими к прямому. Вычисление функции преобразования производится индивидуально для каждой сигнальной плоскости по материалу, набранному за каждые несколько часов, что позволяет эффективно отслеживать вариации $x(t)$ при изменениях настроек питающих напряжений, а также колебаниях температуры и состава газовой смеси. Одним из недостатков изложенного упрощенного подхода является ухудшение точности определения координат для треков, образующих заметный угол с перпендикуляром к плоскости камеры. Рис. 12 показывает зависимость координатной точности одной из сигнальных плоскостей, полученной при проведении треков по срабатываниям в четырех плоскостях вертикальной проекции одного из плеч установки. При рабочем напряжении 3.2 кВ неопределенность восстановления координат составляет приблизительно 200 мкм для перпендикулярных треков. Ее увеличение до 350 мкм для треков, наклоненных на 17°, вполне удовлетворяет требованиям, предъявляемым к точности восстановления кинематики события в установке. Дальнейшее увеличение угла приводит лишь к незначительному ухудшению ситуации. Рост ошибки определения координат с уменьшением питающего напряжения также имеет естественное объяснение. При оптимальной настройке низкий порог дискриминации обеспечивает регистрацию сигнала проволочки от одного-двух первых кластеров заряда на треке. При уменьшении напряжения и, соответственно, газового усиления, для срабатывания дискриминатора при том же пороге необходимо собрать большее количество кластеров заряда, а наступление такого события сильнее разбросано во времени.

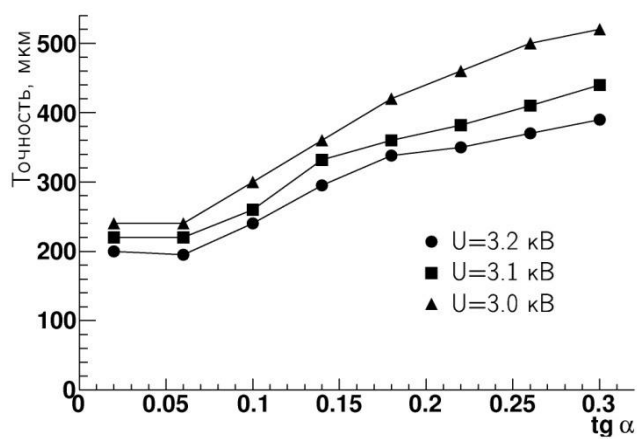

**Рис. 12**. *Зависимость точности дрейфовых камер от тангенса угла наклона трека при различных напряжениях на потенциальных проволочках и фиксированном пороге дискриминации. Угол отсчитывается от нормали к плоскости камеры в плоскости, перпендикулярной направлению проволочек.*

## СИСТЕМА ПОДГОТОВКИ ГАЗОВ

Суммарный газовый объем пропорциональных камер установки невелик и составляет около 10 л, так что при расходе газовой смеси 0.2 л/мин время подготовки камер к работе составляет 8-10 часов, а одного баллона наиболее расходуемого газа – аргона – хватает на 30 суток непрерывного поддержания указанного режима. При отсутствии необходимости изменять скорость продува в больших пределах, для подготовки смеси была выбрана традиционная схема смешивания «по расходу» (рис. 13). Во избежание больших неопределенностей при дозировке сверхмалого расхода фреона-13В1 (необходимое количество 0.15%) используется предварительная подготовка смеси 1.5% $CF_3Br$ 98.5% Ar при давлении около 70 атм в отдельном баллоне. Система смешивания схематично изображена на рис. 13 и имеет три одинаковые ветви дозировки компонентов смеси. Регуляторы давления РД1-3 (РДФ-3), поддерживают постоянное давление 0.12 атм на входах натекателей Н1-3, позволяющих индивидуально регулировать расход каждого компонента смеси. Установка и контроль давления производится с помощью образцовых манометров М1-3, а поток газа в каждой ветви измеряется электронными датчиками расхода ДР1-3 «Honeywell» AWM3100V [23]. Смеситель имеет две выходные ветви, образованные натекателями Н4-5 и датчиками ДР4-5 того же типа, для раздельной регулировки продува блоков камер, расположенных в первом и втором фокусах установки. При сборке системы использованы натекатели и соединительные элементы фирмы «Festo» [31]. Все датчики расхода прокалиброваны на том газе или газовой смеси, поток которой они измеряют. Для получения требуемой газовой смеси устанавливаются следующие соотношения расходов компонент: 65% Ar, 10% Ar-$CF_3Br$, 25% изобутан. Диапазон регулировки выходного потока 0.1−0.5 л/мин.

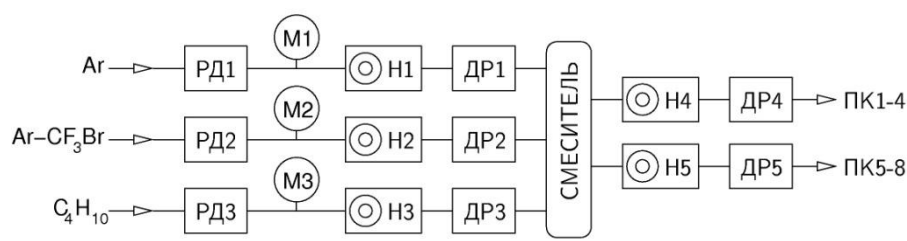

**Рис. 13**. *Схема подготовки газовой смеси для продува пропорциональных камер – смешивание по расходу*

Для местного и дистанционного контроля режима смешивания используется модуль оцифровки E-154 фирмы «L-Card» [22]. Сигналы напряжения с датчиков ДР1-5 преобразуются в значения расходов с помощью калибровочных кривых, заложенных в микропрограмму модуля. Локально расположенный жидкокристаллический дисплей позволяет выводить показания датчиков в форме абсолютного расхода, процентных соотношений или непосредственно оцифрованных напряжений. Загрузка микропрограммы и считывание данных осуществляется по интерфейсу USB1.1. Приложение «Labview» на экране оператора позволяет видеть историю изменения режима смешивания и выдавать сигнал звукового оповещения в случае выхода состава смеси за пределы заданного диапазона.

Дрейфовые камеры установки имеют газовый объем около 650 л, что существенно отличает их от пропорциональных по условиям газового обеспечения. Для приведения камер в рабочее состояние требуется расход газовой смеси около 5 л/мин в течение суток. Поддержание такого режима во все время проведения эксперимента чрезвычайно неэкономично — при длительной работе камер вполне достаточно в несколько раз меньшего продува. Необходимость оперативной перестройки режима газообеспечения потребовала реализации подхода, позволяющего изменять расход газовой смеси в значительных пределах при точном сохранении ее состава. Упрощенная схема системы, удовлетворяющей приведенным требованиям, приведена на рис. 14. Требуемый состав газовой смеси приготавливается в смесителе путем поочередной добавки каждого из компонентов открытием электромагнитных клапанов ЭК1-3 («Технопроект» КЭО 03/10/50/121 [32]). Количество добавленного газа определяется по изменению давления в смесителе объемом около 20 л, измеряемого датчиком ДД1 («Honeywell» ML100PS1DC [23]) с дополнительным контролем по манометру М1. Обратные клапаны ОК1-3 препятствуют обратному току смеси в случаях, когда давление в смесителе превышает давление в баллоне. По достижении расчетного состава смеси открывается клапан ЭК4 и смесь перепускается в буферную емкость объемом 150 л до выравнивания давлений. Скорость расходования смеси определяется установкой натекателя Н1 и варьируется пропорционально изменениям давления в буфере в пределах ±15% от средней величины. При изменении среднего расхода изменяется лишь частота циклов смешивания, тогда как постоянство состава смеси гарантируется самой процедурой дозирования компонентов. Электромагнитный клапан сброса в атмосферу ЭК5 используется для промывки смесителя после длительных перерывов в работе, а также для сброса давления из системы при аварийном неконтролируемом нарастании. Пары вентилей В2-3 и В4-5 обеспечивают смену баллонов быстрорасходуемых газов без нарушения бесперебойной работы системы. В нерабочем состоянии в камерах поддерживается азотная атмосфера путем закрытия В6 и открытия В7 с возможностью регулировки потока азота натекателем Н2 и его индикации по поплавковому ротаметру РТ1.

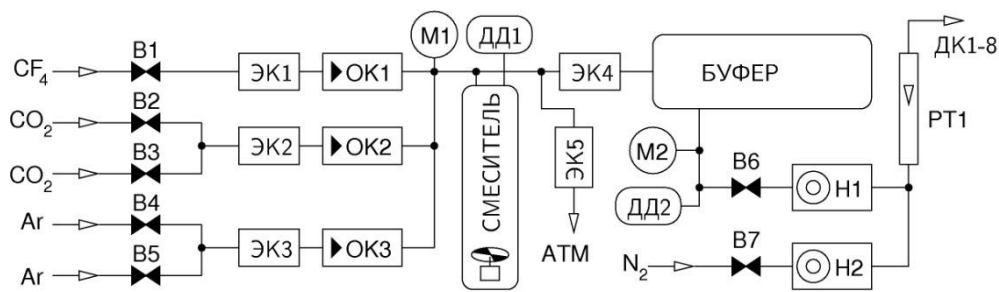

**Рис. 14**. *Схема подготовки газовой смеси для продува дрейфовых камер – смешивание по давлению*

Управление смешиванием осуществляет цифровая плата на основе микроконтроллера PIC16F871 («Microchip» [33]). Когда избыточное давление в буфере, измеряемое датчиком ДД2 (визуальный контроль по манометру М2) опускается ниже 0.7 атм, принимается решение о начале цикла смешивания. Доза каждого компонента смеси рассчитывается алгоритмом исходя из измеренного давления в смесителе, предположения о текущем составе смеси и заданных значений параметров смеси и максимального избыточного давления (3 атм). При расширении газа из баллона в смеситель происходит его охлаждение, поэтому добавление каждого газа производится в три этапа. На первом и втором этапах перепускается такое количество газа, чтобы прибавка давления составила 80% от остатка до расчетной величины, выдерживается пауза в 10 с для установления теплового равновесия и измеряется фактически установившееся давление; малая добавка последнего этапа практически не оказывает влияния на температуру смеси и позволяет установить требуемое значение давления с точностью лучше 0.02 атм. Для ускорения термализации и перемешивания газов в смесителе установлен вентилятор. Микроконтроллер измеряет частоту вращения лопастей при помощи оптического датчика, обеспечивает 30-секундное перемешивание после окончания добавления компонентов и выключает вентилятор в перерывах между циклами смешивания для продления его ресурса работы. Программа микроконтроллера и параметры алгоритма смешивания хранятся во flash-памяти устройства, что обеспечивает возможность работы системы в полностью автономном режиме. При наступлении аварийной ситуации подается звуковой сигнал, по которому можно судить о возникшей неисправности. Предусмотрена возможность ручного управления исполнительными элементами системы, а также визуального контроля показаний датчиков давления на цифровом дисплее. Для удаленного наблюдения имеется возможность передачи показаний датчиков и состояния системы по интерфейсу RS-232, возможно также изменение параметров алгоритма по тому же каналу связи без остановки программы. Приложение «Labview» обеспечивает индикацию динамики важнейших показателей, включая фактический расход смеси, вычисляемый по скорости снижения давления в буфере.

В непосредственной близости от блоков дрейфовых камер расположен распределительный модуль, позволяющий регулировать проток газовой смеси индивидуально для каждой из камер при помощи натекателей. Величина протока по каждой из ветвей измеряется электронными датчиками расхода «Honeywell» и оцифровывается модулем E-154 [22]. Этот же модуль обеспечивает локальную индикацию параметров расхода на жидкокристаллическом дисплее и передачу значений в систему сбора данных для дистанционного контроля.

Описанная система подготовки газовой смеси для дрейфовых камер отлично зарекомендовала себя при проведении эксперимента. Длительность цикла смешивания при приведенных параметрах алгоритма составляет около 2 минут, что позволяет оперативно, регулировкой одного натекателя, изменять расход смеси в пределах 0.5−10 л/мин, при этом периодичность смешивания изменяется от 80 до 4 мин. Планируемое расширение установки предполагает более чем тройное увеличение газового объема дрейфовых камер, соответственно должен быть увеличен и их продув. Необходимого для подготовки камер расхода 15−20 л/мин несложно достичь без изменения конструкции, достаточно увеличить параметр максимального давления смешивания до 5−7 атм. Предполагается дополнительно оснастить систему датчиками давления в баллонах для своевременного предупреждения оператора о необходимости замены.

# ТРИГГЕРНАЯ СИСТЕМА

При наборе физической статистики в установке ЭПЕКУР используется тщательно продуманная логика выработки сигналов запуска для записи событий. При формировании различных логических комбинаций используются сцинтилляционные счетчики С1, С2, А (рис. 3), а также быстрые сигналы от сборок пропорциональных камер в первом и втором фокусах канала ПК1-4 и ПК5-8. Последние, в свою очередь, вырабатываются в специальных распределительных блоках (см. [28]) как результат мажоритарной логики, требующей срабатывания хотя бы 3 из 8 плоскостей камер, входящих в каждую сборку. Эти же распределительные блоки ответственны за распространение результирующего триггерного сигнала установки на каждую из плат электроники считывания. Основной физический триггер формируется при условии прохождения частицы пучка $П = С1 \cdot ПК1\text{-}4 \cdot ПК5\text{-}8$ через выделяющий счетчик мишени С2 и ее выбывании из пучка после прохождения мишени: $Т_0 = П \cdot С2 \cdot \overline{А}$. Даже при таком мягком условии вырабатывается всего несколько тысяч событий основного триггера $Т_0$ за сброс ускорителя, тогда как система считывания способна без потерь обрабатывать до 50 тысяч событий за период повторения циклов. Логическая комбинация $Т_1 = П \cdot С2$ содержит только условие прохождения пучковой частицы через мишень и, таким образом, не подвержена влиянию хода физического сечения. Триггерный сигнал по этому условию, подвергнутый 100-кратному прореживанию, используется для нормировки данных на количество частиц пучка, прошедших через мишень. Программы обработки используют одинаковые алгоритмы для событий $Т_0$ и $Т_1$ при восстановлении треков в блоках пучковых камер и определении импульса частиц пучка, что позволяет исключить влияние неэффективности как самих камер, так и алгоритмов восстановления на процедуру нормировки в любом импульсном интервале. Вспомогательный триггер $Т_2 = П \cdot А$, также прореженный в 100 раз, используется для получения пространственного распределения частиц пучка на мишени, неискаженного малым размером счетчика С2. Положение максимума этого распределения используется для прецизионного контроля полей поворотных магнитов П3 и П4, задающих центральный импульс канала.

Триггерные сигналы установки ЭПЕКУР формируются в модуле, использующем микросхему программируемой логики фирмы Xilinx XC3S400 [34]. В основе модуля лежит модифицированная плата считывания с пропорциональных камер [28], дополненная преобразователями уровней NIM/CMOS для входных сигналов и CMOS/NIM для выходных. Загрузка конфигурации ПЛИС и управление ее работой производится со стороны USB2.0 интерфейса при помощи микропроцессора CY7C68013A, расположенного на той же плате. На рис. 15 приведена сильно упрощенная блок-схема модуля -- проиллюстрирована лишь та часть конфигурации ПЛИС, которая отвечает непосредственно за логику выработки основных триггерных сигналов. Импульсы от счетчиков С1, С2, А формируются дискриминаторами с фиксированным порогом ДФП и попадают на входы логической матрицы параллельно с сигналами ПК1-4 и ПК5-8. Взаимная синхронизация сигналов обеспечивается линиями задержки ЛЗ и цифровой настройкой выходов распределительных блоков. Особенностью ПЛИС семейства FPGA является строгая необходимость тактирования внутренних сигналов. В реализации модуля для тактирования используется сигнал от счетчика С1, оптимизированного по временному разрешению и участвующему во всех триггерных комбинациях. Передний фронт сигнала С1 устанавливает триггер Т0, что приводит к одновременному защелкиванию состояния остальных входных сигналов в триггерах Т1-Т8. Каждая из схем совпадений СС1-СС6 может быть индивидуально сконфигурирована для выполнения операции логического умножения некоторых из сигналов Т1-Т8 непосредственно или с инверсией, позволяя таким образом реализовывать одновременно до 6 логических комбинаций входных сигналов (физических триггеров). Программируемые счетчики ПС1-ПС6 могут обеспечивать прореживание физических триггеров с кратностью до $2^{16}$. Каждый из физических триггеров может приводить к выработке одного или нескольких сигналов запуска определенных частей установки в выходных формирователях ВФ1-ВФ3. В рабочей конфигурации при наступлении события основного триггера $T_0$ запускаются для считывания пропорциональные (ПК) и дрейфовые (ДК) камеры, тогда как события $T_1$ и $T_2$ вызывают только обработку пропорциональных камер ПК. Третий выход (ВПС) предполагалось задействовать для измерения времени пролета частиц пучка, однако в настоящее время эта подсистема не используется. Конфигурация выходных формирователей ВФ1-ВФ3 позволяет программировать длительность выходного сигнала от одного до 16 периодов частоты 192 МГц, а также блокировать последующую выработку физических триггеров на время до 4096 периодов той же частоты. Выработка выходных сигналов запуска производится через 7 нс после прихода тактирующего сигнала С1, тогда же сбрасывается триггер Т0, приводя систему в состояние готовности. Важно, что за это время логическое состояние Т1-Т8 успевает распространиться через всю внутреннюю логику ПЛИС, а сигналы на входах ВФ1-ВФ3 устанавливаются в определенные значения. Таким образом, примененный подход позволяет соблюсти строгие требования тактирования сигналов ПЛИС, а также обеспечить временную привязку выходных сигналов запуска к моменту срабатывания счетчика С1.

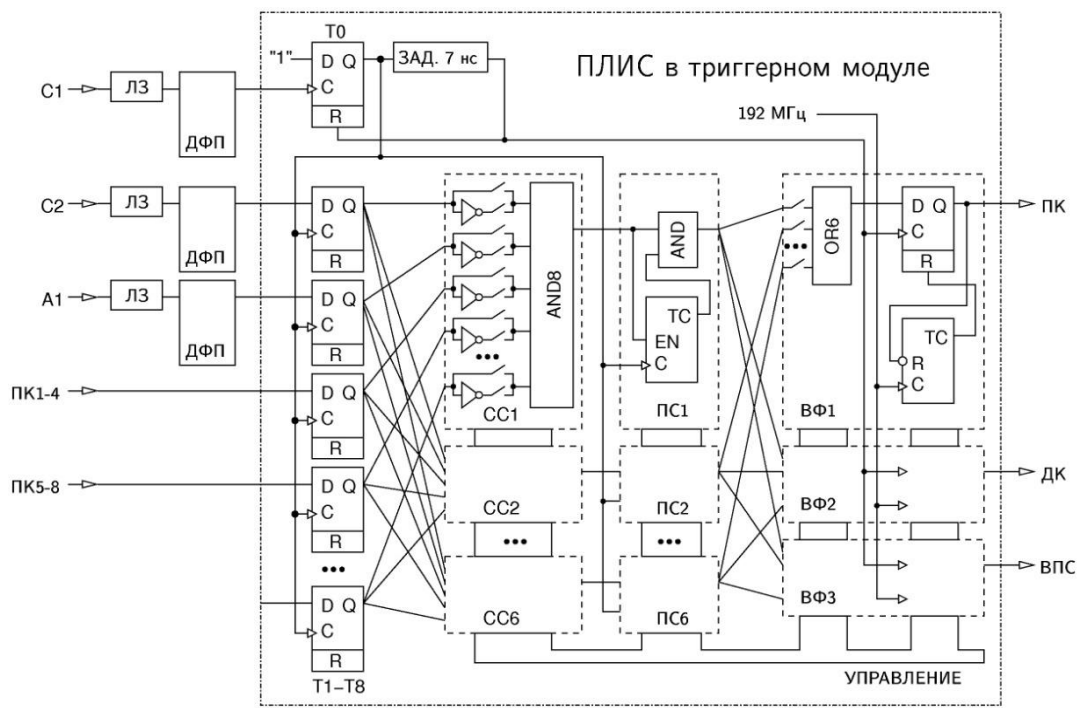

**Рис. 15**. *Упрощенная блок-схема выработки триггера*

Использование ПЛИС в триггерном модуле открывает множество дополнительных возможностей, таких как подсчет количества различных импульсов, включая индивидуальные входы, физические триггера и фактически выработанные запуски. Избыточность конфигурации позволяет гибко настраивать модуль и при наладочных работах: незадействованные на рис. 15 входы используются для дополнительных счетчиков, а свободные схемы совпадений – для получения отладочных логических комбинаций. Программирование модуля, включая задание логики физических триггеров, осуществляется с помощью приложения «Labview», оно же позволяет наблюдать значения и историю различных счетов и их отношений, а также передавать их на центральный сервер системы сбора данных для записи в поток.

## АВТОМАТИЗАЦИЯ ИЗМЕРЕНИЯ ПОЛЯ ПО ЯМР

Прецизионный характер настройки магнитного канала на заданный импульс пучка требует не просто однократного измерения поля поворотного магнита Б322С2 с помощью ЯМР, но и непрерывного контроля его величины во время набора статистики. Используемые для получения ЯМР-сигнала стандартные приборы Ш1-1 или Ш1-9 подвергнуты незначительной модификации в части вывода сигналов синусоидальной модуляции поля, частоты возбуждения и собственно выпрямленного сигнала ЯМР-поглощения таким образом, чтобы обеспечить возможность их передачи по коаксиальным кабелям на расстояние до 30 м. Расположенный вблизи операторского компьютера модуль «L-Card» E-154 [22] производит оцифровку сигналов поглощения и модуляции с частотой 10 кГц, а также измерение частоты возбуждения путем сопоставления с эталонным кварцевым генератором 24.0 МГц. Точность измерения частоты не хуже 0.05% обеспечивается требованиями стандарта USB, для которой используется тот же кварцевый генератор, и может быть улучшена до 0.01% после калибровки по промышленному частотомеру. Компьютерная обработка оцифрованных сигналов позволяет не только определить величину поля при точной настройке частоты в резонанс, но и, после предварительной калибровки, пересчитывать точное значение поля в случаях небольшого отклонения частоты от резонансного значения с учетом измеренной амплитуды модуляции и дисбаланса сигналов резонансного поглощения. Предусмотрена также возможность подстройки частоты возбуждения путем добавления уровня с выхода ЦАП E-154 к напряжению управления варикапом контура в приборах Ш1-1 или Ш1-9, однако текущими алгоритмами эта возможность не используется. Приложение «Labview» обеспечивает визуализацию оцифрованных сигналов и полученных величин поля, а также их передачу на сервер системы сбора данных для включения в поток записываемой информации.

# ЗАКЛЮЧЕНИЕ

Идея эксперимента ЭПЕКУР возникла в апреле 2004 года и к осени того же года получила одобрение научного сообщества обоих российских институтов, составивших основу коллаборации. Чрезвычайно ограниченное финансирование программы началось в мае 2005 года. В условиях нехватки средств и квалифицированного персонала, коллективу коллаборации понадобилось всего три с половиной года для того, чтобы на пустом месте создать действующую установку -- в декабре 2008 года в рамках первого инженерного сеанса ЭПЕКУР записал первые события пион-протонного рассеяния. Четыре двухнедельных сеанса набора статистики, проведенных в течение трех последующих лет, позволили накопить около 3 миллиардов триггерных событий. Обработка записанного материала привела к появлению нового поколения экспериментальных данных по дифференциальному сечению упругого пион-протонного рассеяния, а также к обнаружению узкой особенности в энергетическом ходе сечения при инвариантной массе системы 1685 МэВ. К большому сожалению команды ЭПЕКУР, пожар, произошедший на ускорительном комплексе ИТЭФ в феврале 2012 года, нарушил планы коллаборации по дальнейшему исследованию узкой особенности и расширению программы прецизионного измерения сечений.

Несмотря на все трудности, которые пришлось преодолеть членам коллаборации для достижения успеха, ЭПЕКУР стал едва ли не единственной экспериментальной установкой подобного масштаба в России, созданной в кратчайшие сроки «с нуля» в непростые постперестроечные времена. Методические подходы, использованные при конструировании отдельных подсистем, хотя и не представляют собой серьезного шага вперед в области детектирования частиц или построения экспериментальных установок, являются обобщением огромного опыта участников коллаборации и результатом глубоко продуманной оптимизации инженерных решений в условиях жесткого ограничения ресурсов. Оснащение традиционных детекторов самой современной электроникой, использующей, в том числе, высокую плотность расположения элементов и производительность устройств на основе программируемых логических матриц и микропроцессоров, позволило обеспечить детектирующей системе высокую эффективность регистрации, надежность и стабильность поддержания режимов. Значительный запас по скорости сбора данных обеспечивается применением высокоскоростных интерфейсов USB2.0 и Ethernet. Благодаря использованию микропроцессорных элементов и современных дистанционных измерительных средств во вспомогательных подсистемах, весь контроль за параметрами может быть сосредоточен на рабочем месте единственного оператора, что существенно облегчает эксплуатацию комплекса. Таким образом, созданная при минимальных затратах ресурсов экспериментальная установка ЭПЕКУР отвечает самым современным требованиям и оптимальна для решения поставленных физических задач.

# СПИСОК ЛИТЕРАТУРЫ